\documentclass[10pt,a4paper,onecolumn]{article}
\usepackage{marginnote}
\usepackage{graphicx}
\usepackage{xcolor}
\usepackage{authblk,etoolbox}
\usepackage{titlesec}
\usepackage{calc}
\usepackage{tikz}
\usepackage{hyperref}
\hypersetup{colorlinks,breaklinks=true,
            urlcolor=[rgb]{0.0, 0.5, 1.0},
            linkcolor=[rgb]{0.0, 0.5, 1.0}}
\usepackage{caption}
\usepackage{tcolorbox}
\usepackage{amssymb,amsmath}
\usepackage{ifxetex,ifluatex}
\usepackage{seqsplit}
\usepackage{xstring}

\usepackage{float}
\let\origfigure\figure
\let\endorigfigure\endfigure
\renewenvironment{figure}[1][2] {
    \expandafter\origfigure\expandafter[H]
} {
    \endorigfigure
}

\usepackage{fixltx2e} 
\usepackage[
  backend=biber,
]{biblatex}
\bibliography{paper.bib}

\let\textttOrig=\texttt
\def\texttt#1{\expandafter\textttOrig{\seqsplit{#1}}}
\renewcommand{\seqinsert}{\ifmmode
  \allowbreak
  \else\penalty6000\hspace{0pt plus 0.02em}\fi}

\makeatletter
\let\href@Orig=\href
\def\href@Urllike#1#2{\href@Orig{#1}{\begingroup
    \def\Url@String{#2}\Url@FormatString
    \endgroup}}
\def\href@Notdoi#1#2{\def\tempa{#1}\def\tempb{#2}%
  \ifx\tempa\tempb\relax\href@Urllike{#1}{#2}\else
  \href@Orig{#1}{#2}\fi}
\def\href#1#2{%
  \IfBeginWith{#1}{https://doi.org}%
  {\href@Urllike{#1}{#2}}{\href@Notdoi{#1}{#2}}}
\makeatother

\newlength{\cslhangindent}
\newlength{\csllabelwidth}
\newenvironment{CSLReferences}[3] 
 {
  \setlength{\parindent}{0pt}
  \ifodd #1 \everypar{\setlength{\hangindent}{\cslhangindent}}\ignorespaces\fi
  \ifnum #2 > 0
  \setlength{\parskip}{#2\baselineskip}
  \fi
 }%
 {}
\usepackage{calc}

\usepackage[top=3.5cm, bottom=3cm, right=1.5cm, left=1.0cm,
            headheight=2.2cm, reversemp, includemp, marginparwidth=4.5cm]{geometry}

\titleformat{\section}
  {\normalfont\sffamily\Large\bfseries}
  {}{0pt}{}
\titleformat{\subsection}
  {\normalfont\sffamily\large\bfseries}
  {}{0pt}{}
\titleformat{\subsubsection}
  {\normalfont\sffamily\bfseries}
  {}{0pt}{}
\titleformat*{\paragraph}
  {\sffamily\normalsize}

\usepackage{fancyhdr}
\makeatletter
\let\ps@plain\ps@fancy
\fancyheadoffset[L]{4.5cm}
\fancyfootoffset[L]{4.5cm}

\definecolor{linky}{rgb}{0.0, 0.5, 1.0}

\newtcolorbox{repobox}
   {colback=red, colframe=red!75!black,
     boxrule=0.5pt, arc=2pt, left=6pt, right=6pt, top=3pt, bottom=3pt}

\newcommand{\ExternalLink}{%
   \tikz[x=1.2ex, y=1.2ex, baseline=-0.05ex]{%
       \begin{scope}[x=1ex, y=1ex]
           \clip (-0.1,-0.1)
               --++ (-0, 1.2)
               --++ (0.6, 0)
               --++ (0, -0.6)
               --++ (0.6, 0)
               --++ (0, -1);
           \path[draw,
               line width = 0.5,
               rounded corners=0.5]
               (0,0) rectangle (1,1);
       \end{scope}
       \path[draw, line width = 0.5] (0.5, 0.5)
           -- (1, 1);
       \path[draw, line width = 0.5] (0.6, 1)
           -- (1, 1) -- (1, 0.6);
       }
   }

\patchcmd{\@maketitle}{center}{flushleft}{}{}
\patchcmd{\@maketitle}{center}{flushleft}{}{}
\patchcmd{\@maketitle}{\LARGE}{\LARGE\sffamily}{}{}
\def\maketitle{{%
  
  \AB@maketitle}}
\makeatletter
\renewcommand\AB@affilsepx{ \protect\Affilfont}
\renewcommand\AB@affilnote[1]{{\bfseries #1}\hspace{3pt}}
\renewcommand{\affil}[2][]%
   {\newaffiltrue\let\AB@blk@and\AB@pand
      \if\relax#1\relax\def\AB@note{\AB@thenote}\else\def\AB@note{#1}%
        \setcounter{Maxaffil}{0}\fi
        \begingroup
        \let\href=\href@Orig
        \let\texttt=\textttOrig
        \let\protect\@unexpandable@protect
        \def\thanks{\protect\thanks}\def\footnote{\protect\footnote}%
        \@temptokena=\expandafter{\AB@authors}%
        {\def\\{\protect\\\protect\Affilfont}\xdef\AB@temp{#2}}%
         \xdef\AB@authors{\the\@temptokena\AB@las\AB@au@str
         \protect\\[\affilsep]\protect\Affilfont\AB@temp}%
         \gdef\AB@las{}\gdef\AB@au@str{}%
        {\def\\{, \ignorespaces}\xdef\AB@temp{#2}}%
        \@temptokena=\expandafter{\AB@affillist}%
        \xdef\AB@affillist{\the\@temptokena \AB@affilsep
          \AB@affilnote{\AB@note}\protect\Affilfont\AB@temp}%
      \endgroup
       \let\AB@affilsep\AB@affilsepx
}
\makeatother

\renewcommand\Affilfont{\sffamily\small\mdseries}
\ifnum 0\ifxetex 1\fi\ifluatex 1\fi=0 
  \usepackage[T1]{fontenc}
  \usepackage[utf8]{inputenc}

\else 
  \ifxetex
    \usepackage{mathspec}
    \usepackage{fontspec}

  \else
    \usepackage{fontspec}
  \fi
  \defaultfontfeatures{Ligatures=TeX,Scale=MatchLowercase}

\fi
\IfFileExists{upquote.sty}{\usepackage{upquote}}{}
\IfFileExists{microtype.sty}{%
\usepackage{microtype}
\UseMicrotypeSet[protrusion]{basicmath} 
}{}

\usepackage{hyperref}
\hypersetup{unicode=true,
            pdftitle={pySYD: Automated measurements of global asteroseismic parameters},
            pdfborder={0 0 0},
            breaklinks=true}
\let\addcontentslineOrig=\addcontentsline
\def\addcontentsline#1#2#3{\bgroup
  \let\texttt=\textttOrig\addcontentslineOrig{#1}{#2}{#3}\egroup}
\let\markbothOrig\markboth
\def\markboth#1#2{\bgroup
  \let\texttt=\textttOrig\markbothOrig{#1}{#2}\egroup}
\let\markrightOrig\markright
\def\markright#1{\bgroup
  \let\texttt=\textttOrig\markrightOrig{#1}\egroup}

\usepackage{graphicx,grffile}
\makeatletter
\def\maxwidth{\ifdim\Gin@nat@width>\linewidth\linewidth\else\Gin@nat@width\fi}
\def\maxheight{\ifdim\Gin@nat@height>\textheight\textheight\else\Gin@nat@height\fi}
\makeatother
\setkeys{Gin}{width=\maxwidth,height=\maxheight,keepaspectratio}
\IfFileExists{parskip.sty}{%
\usepackage{parskip}
}{
\setlength{\parindent}{0pt}
\setlength{\parskip}{6pt plus 2pt minus 1pt}
}
\providecommand{\tightlist}{%
  \setlength{\itemsep}{0pt}\setlength{\parskip}{0pt}}
\ifx\paragraph\undefined\else
\let\oldparagraph\paragraph
\renewcommand{\paragraph}[1]{\oldparagraph{#1}\mbox{}}
\fi
\ifx\subparagraph\undefined\else
\let\oldsubparagraph\subparagraph
\renewcommand{\subparagraph}[1]{\oldsubparagraph{#1}\mbox{}}
\fi

\title{\texttt{pySYD}: Automated measurements of global asteroseismic
parameters}

        \author[1, 2]{Ashley Chontos}
          \author[1]{Daniel Huber}
          \author[3]{Maryum Sayeed}
          \author[4]{Pavadol Yamsiri}
    
      \affil[1]{Institute for Astronomy, University of Hawai'i, 2680
Woodlawn Drive, Honolulu, HI 96822, USA}
      \affil[2]{NSF Graduate Research Fellow}
      \affil[3]{Department of Astronomy, Columbia University, Pupin Physics Laboratories, New York, NY 10027, USA}
      \affil[4]{Sydney Institute for Astronomy, School of Physics,
University of Sydney, NSW 2006, Australia}
  \date{\vspace{-7ex}}

\begin{document}
\maketitle

\marginpar{

  \begin{flushleft}
  \sffamily\small

  {\bfseries DOI:} \href{https://doi.org/DOI unavailable}{\color{linky}{DOI unavailable}}

  \vspace{2mm}

  {\bfseries Software}
  \begin{itemize}
    \setlength\itemsep{0em}
    \item \href{N/A}{\color{linky}{Review}} \ExternalLink
    \item \href{NO_REPOSITORY}{\color{linky}{Repository}} \ExternalLink
    \item \href{DOI unavailable}{\color{linky}{Archive}} \ExternalLink
  \end{itemize}

  \vspace{2mm}

  \par\noindent\hrulefill\par

  \vspace{2mm}

  {\bfseries Editor:} \href{https://example.com}{Pending
Editor} \ExternalLink \\
  \vspace{1mm}
    {\bfseries Reviewers:}
  \begin{itemize}
  \setlength\itemsep{0em}
    \item \href{https://github.com/Pending Reviewers}{@Pending
Reviewers}
    \end{itemize}
    \vspace{2mm}

  {\bfseries Submitted:} N/A\\
  {\bfseries Published:} N/A

  \vspace{2mm}
  {\bfseries License}\\
  Authors of papers retain copyright and release the work under a Creative Commons Attribution 4.0 International License (\href{http://creativecommons.org/licenses/by/4.0/}{\color{linky}{CC BY 4.0}}).

  \end{flushleft}
}

\hypertarget{summary}{%
\section{Summary}\label{summary}}

Asteroseismology, the study of stellar oscillations, is a powerful tool
for studying the interiors of stars and determining their fundamental
properties (Aerts, 2021). For stars with temperatures that are similar
to the Sun, turbulent near-surface convection excites sound waves that
propagate within the stellar cavity (Bedding, 2014). These waves
penetrate into different depths within the star and therefore provide
powerful constraints on stellar interiors that would otherwise be
inaccessible. Asteroseismology is well-established in astronomy as the
gold standard for characterizing fundamental properties like masses,
radii, densities, and ages for single stars, which has broad impacts on
several fields in astronomy. For example, ages of stars are important to
reconstruct the formation history of the Milky Way (so-called galactic
archaeology). For exoplanets that are discovered indirectly through
changes in stellar observables, precise and accurate stellar masses and
radii are critical for learning about the planets that orbit them.

\hypertarget{statement-of-need}{%
\section{Statement of Need}\label{statement-of-need}}

The NASA space telescopes \emph{Kepler}, K2 and TESS have recently
provided very large databases of high-precision light curves of stars.
By detecting brightness variations due to stellar oscillations, these
light curves allow the application of asteroseismology to large numbers
of stars, which requires automated software tools to efficiently extract
observables. Several tools have been developed for asteroseismic
analyses (e.g., \texttt{A2Z}, Mathur et al., 2010; \texttt{COR}, Mosser
\& Appourchaux, 2009; \texttt{OCT}, Hekker et al., 2010; \texttt{SYD},
Huber et al., 2009), but many of them are closed-source and therefore
inaccessible to the general astronomy community. Some open-source tools
exist (e.g., \texttt{DIAMONDS} and \texttt{FAMED}, Corsaro \& De Ridder,
2014; \texttt{PBjam}, Nielsen et al., 2021; \texttt{lightkurve},
Lightkurve Collaboration et al., 2018), but they are either optimized
for smaller samples of stars or have not yet been extensively tested
against closed-source tools.

\texttt{pySYD} is adapted from the framework of the IDL-based
\texttt{SYD} pipeline (Huber et al., 2009; hereafter referred to as
\texttt{SYD}), which has been used frequently to measure asteroseismic
parameters for \emph{Kepler} stars and has been extensively tested
against other closed-source tools on \emph{Kepler} data (Hekker et al., 2011;
Verner et al., 2011). Papers based on asteroseismic parameters measured
using the \texttt{SYD} pipeline include Huber et al. (2011), Bastien et
al. (2013), Chaplin et al. (2014), Serenelli et al. (2017), and Yu et
al. (2018). \texttt{pySYD} was developed using the same well-tested
methodology, but has improved functionality including automated
background model selection and parallel processing as well as improved
flexibility through a user-friendly interface, while still maintaining
its speed and efficiency. Well-documented, open-source asteroseismology
software that has been benchmarked against closed-source tools are
critical to ensure the reproducibility of legacy results from the
\emph{Kepler} mission (Borucki et al., 2010). The combination of
well-tested methodology, improved flexibility and parallel processing
capabilities will make \texttt{pySYD} a promising tool for the broader
community to analyze current and forthcoming data from the NASA TESS
mission (Ricker et al., 2015).

\hypertarget{the-pysyd-library}{%
\section{\texorpdfstring{The \texttt{pySYD}
library}{The pySYD library}}\label{the-pysyd-library}}

The excitation mechanism for solar-like oscillations is stochastic and
modes are observed over a range of frequencies. Oscillation modes are
separated by the so-called large frequency spacing (\(\Delta\nu\)) with
an approximately Gaussian-shaped power excess centered on the frequency 
of maximum power (\(\rm \nu_{max}\)). The observables \(\rm \nu_{max}\) 
and \(\Delta\nu\) are directly related to fundamental properties such as 
surface gravity, density, mass and radius (Kjeldsen
\& Bedding, 1995).

\texttt{pySYD} is a Python package for detecting solar-like oscillations
and measuring global asteroseismic parameters. Derived parameters
include \(\rm \nu_{max}\) and \(\Delta\nu\), as well as characteristic
amplitudes and time scales of various granulation processes.

A \texttt{pySYD} pipeline \texttt{Target} class object has two main
methods:

\begin{itemize}
\tightlist
\item
  The first module searches for signatures of solar-like oscillations by
  implementing a frequency-resolved, collapsed autocorrelation (ACF)
  method. The output from this routine provides an estimate for
  \(\rm \nu_{max}\), which is used as an initial guess for the main
  pipeline function (i.e.~second module). However if \(\rm \nu_{max}\)
  is already known, the user can provide an estimate and hence bypass
  this module.
\item
  The second routine begins by masking out the region in the power
  spectrum (PS) with the power excess in order to characterize the
  stellar background. \texttt{pySYD} optimizes and selects the
  stellar background model that minimizes the BIC (Bayesian Information
  Criterion; Schwarz, 1978). The best-fit background model is
  then subtracted from the PS, where the peak of the smoothed,
  background-corrected PS is \(\rm \nu_{max}\). An ACF is computed using
  the region in the power spectrum centered on \(\rm \nu_{max}\) and the
  peak in the ACF closest to the expected value for the large frequency
  separation is adopted as \(\Delta\nu\).
\end{itemize}

\begin{figure}
\centering
\includegraphics{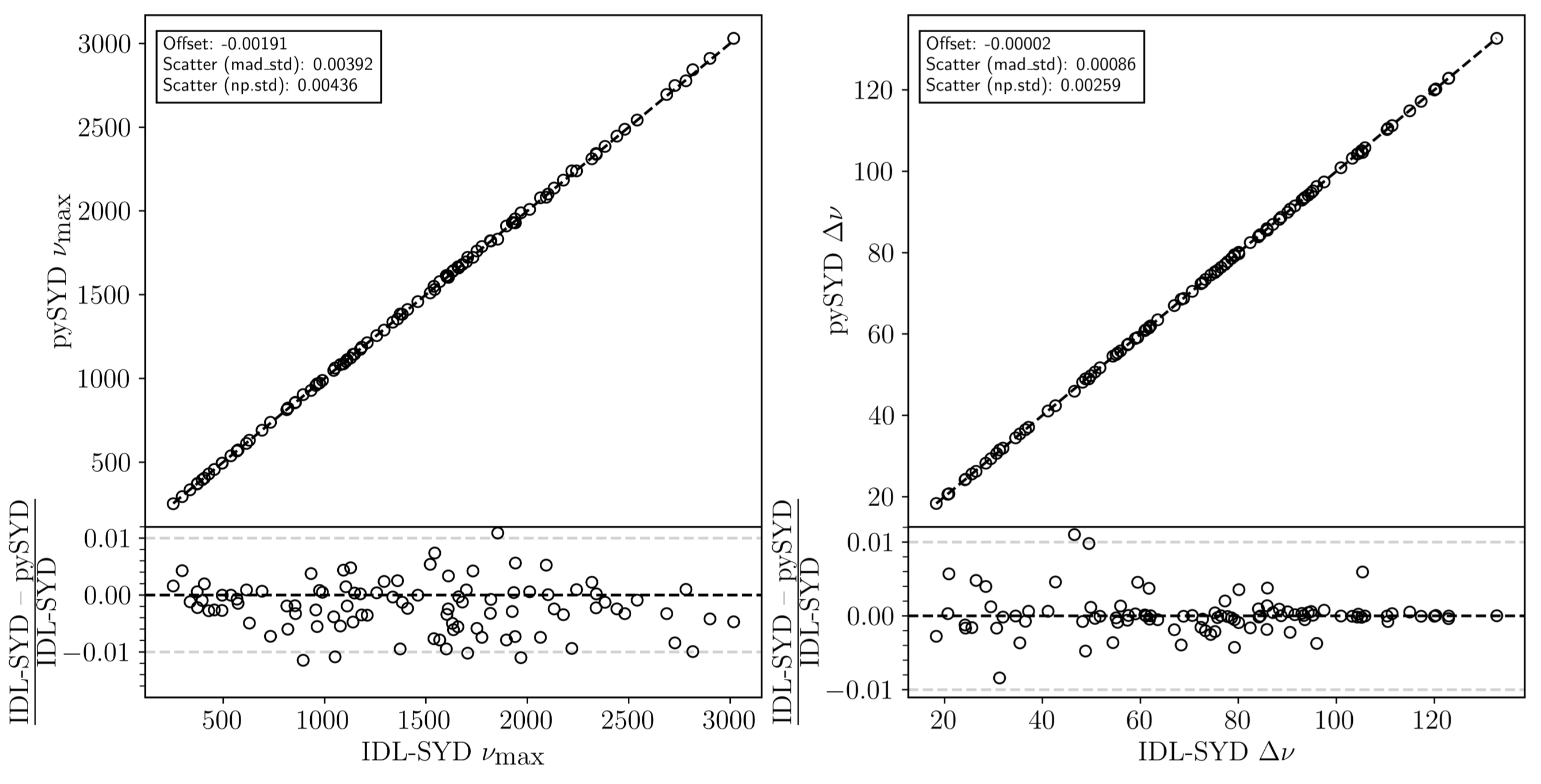}
\caption{Comparison of global parameters \(\rm \nu_{max}\) (left) and
\(\Delta\nu\) (right) measured by \texttt{pySYD} and IDL \texttt{SYD}
for \(\sim100\) \emph{Kepler} stars. The bottom panels show the
fractional residuals.\label{fig:comparison}}
\end{figure}

The \texttt{pySYD} software depends on a number of powerful libraries,
including Astropy (Astropy Collaboration et al., 2018, 2013), Matplotlib
(Hunter, 2007), Numpy (Harris et al., 2020), and SciPy (Virtanen et al.,
2020). \texttt{pySYD} has been tested against \texttt{SYD} using results
from the \emph{Kepler} sample for \(\sim100\) stars
(\autoref{fig:comparison}). The comparisons show no significant
systematic differences, with a median offset and scatter of \(0.2\%\)
and \(0.4\%\) for \(\rm \nu_{max}\) as well as \(0.002\%\) and
\(0.09\%\) for \(\Delta\nu\), which is smaller or comparable to the
typical random uncertainties (Huber et al., 2011).

\hypertarget{documentation-examples}{%
\section{Documentation \& Examples}\label{documentation-examples}}

The main documentation for the \texttt{pySYD} software is hosted at
\href{https://pysyd.readthedocs.io}{pysyd.readthedocs.io}.
\texttt{pySYD} provides a convenient setup feature that will download
data for three example stars and automatically create the recommended
data structure for an easy quickstart. The features of the
\texttt{pySYD} output results are described in detail in the
documentation.

\hypertarget{acknowledgements}{%
\section{Acknowledgements}\label{acknowledgements}}

We thank Dennis Stello, Jie Yu, Marc Hon, Yifan Chen, Yaguang Li, and 
other \texttt{pySYD} users for discussion and suggestions which helped 
with the development of this code.

We also acknowledge support from:
\begin{itemize}
\tightlist
\item 
  The National Science Foundation (DGE-1842402, AST-1717000)
\item 
  The National Aeronautics and Space Administration (80NSSC19K0597)
\item
  The Alfred P. Sloan Foundation
\end{itemize}

\hypertarget{references}{%
\section*{References}\label{references}}
\addcontentsline{toc}{section}{References}

\hypertarget{refs}{}
\begin{CSLReferences}{1}{0}
\leavevmode\hypertarget{ref-aerts2021}{}%
Aerts, C. (2021). {Probing the interior physics of stars through
asteroseismology}. \emph{Reviews of Modern Physics}, \emph{93}(1),
015001. \url{https://doi.org/10.1103/RevModPhys.93.015001}

\leavevmode\hypertarget{ref-astropy2}{}%
Astropy Collaboration, Price-Whelan, A. M., Sipőcz, B. M., Günther, H.
M., Lim, P. L., Crawford, S. M., Conseil, S., Shupe, D. L., Craig, M.
W., Dencheva, N., Ginsburg, A., VanderPlas, J. T., Bradley, L. D.,
Pérez-Suárez, D., de Val-Borro, M., Aldcroft, T. L., Cruz, K. L.,
Robitaille, T. P., Tollerud, E. J., \ldots{} Astropy Contributors.
(2018). {The Astropy Project: Building an Open-science Project and
Status of the v2.0 Core Package}. \emph{156}(3), 123.
\url{https://doi.org/10.3847/1538-3881/aabc4f}

\leavevmode\hypertarget{ref-astropy1}{}%
Astropy Collaboration, Robitaille, T. P., Tollerud, E. J., Greenfield,
P., Droettboom, M., Bray, E., Aldcroft, T., Davis, M., Ginsburg, A.,
Price-Whelan, A. M., Kerzendorf, W. E., Conley, A., Crighton, N.,
Barbary, K., Muna, D., Ferguson, H., Grollier, F., Parikh, M. M., Nair,
P. H., \ldots{} Streicher, O. (2013). {Astropy: A community Python
package for astronomy}. \emph{558}, A33.
\url{https://doi.org/10.1051/0004-6361/201322068}

\leavevmode\hypertarget{ref-bastien2013}{}%
Bastien, F. A., Stassun, K. G., Basri, G., \& Pepper, J. (2013). {An
observational correlation between stellar brightness variations and
surface gravity}. \emph{500}(7463), 427--430.
\url{https://doi.org/10.1038/nature12419}

\leavevmode\hypertarget{ref-bedding2014}{}%
Bedding, T. R. (2014). {Solar-like oscillations: An observational
perspective}. In P. L. Pallé \& C. Esteban (Eds.),
\emph{Asteroseismology} (p. 60).

\leavevmode\hypertarget{ref-borucki2010}{}%
Borucki, W. J., Koch, D., Basri, G., Batalha, N., Brown, T., Caldwell,
D., Caldwell, J., Christensen-Dalsgaard, J., Cochran, W. D., DeVore, E.,
Dunham, E. W., Dupree, A. K., Gautier, T. N., Geary, J. C., Gilliland,
R., Gould, A., Howell, S. B., Jenkins, J. M., Kondo, Y., \ldots{} Prsa,
A. (2010). {Kepler Planet-Detection Mission: Introduction and First
Results}. \emph{Science}, \emph{327}, 977.
\url{https://doi.org/10.1126/science.1185402}

\leavevmode\hypertarget{ref-chaplin2014}{}%
Chaplin, W. J., Basu, S., Huber, D., Serenelli, A., Casagrande, L.,
Silva Aguirre, V., Ball, W. H., Creevey, O. L., Gizon, L., Handberg, R.,
Karoff, C., Lutz, R., Marques, J. P., Miglio, A., Stello, D., Suran, M.
D., Pricopi, D., Metcalfe, T. S., Monteiro, M. J. P. F. G., \ldots{}
Salabert, D. (2014). {Asteroseismic Fundamental Properties of Solar-type
Stars Observed by the NASA Kepler Mission}. \emph{210}, 1.
\url{https://doi.org/10.1088/0067-0049/210/1/1}

\leavevmode\hypertarget{ref-corsaro2014}{}%
Corsaro, E., \& De Ridder, J. (2014). {DIAMONDS: A new Bayesian nested
sampling tool. Application to peak bagging of solar-like oscillations}.
\emph{571}, A71. \url{https://doi.org/10.1051/0004-6361/201424181}

\leavevmode\hypertarget{ref-numpy}{}%
Harris, C. R., Millman, K. J., Walt, S. J. van der, Gommers, R.,
Virtanen, P., Cournapeau, D., Wieser, E., Taylor, J., Berg, S., Smith,
N. J., Kern, R., Picus, M., Hoyer, S., Kerkwijk, M. H. van, Brett, M.,
Haldane, A., Río, J. F. del, Wiebe, M., Peterson, P., \ldots{} Oliphant,
T. E. (2020). Array programming with {NumPy}. \emph{Nature},
\emph{585}(7825), 357--362.
\url{https://doi.org/10.1038/s41586-020-2649-2}

\leavevmode\hypertarget{ref-hekker2010}{}%
Hekker, S., Broomhall, A.-M., Chaplin, W. J., Elsworth, Y. P., Fletcher,
S. T., New, R., Arentoft, T., Quirion, P.-O., \& Kjeldsen, H. (2010).
{The Octave (Birmingham-Sheffield Hallam) automated pipeline for
extracting oscillation parameters of solar-like main-sequence stars}.
\emph{402}(3), 2049--2059.
\url{https://doi.org/10.1111/j.1365-2966.2009.16030.x}

\leavevmode\hypertarget{ref-hekker2011}{}%
Hekker, S., Elsworth, Y., De Ridder, J., Mosser, B., Garc\'ia, R. A.,
Kallinger, T., Mathur, S., Huber, D., Buzasi, D. L., Preston, H. L.,
Hale, S. J., Ballot, J., Chaplin, W. J., Régulo, C., Bedding, T. R.,
Stello, D., Borucki, W. J., Koch, D. G., Jenkins, J., \ldots{}
Christensen-Dalsgaard, J. (2011). {Solar-like oscillations in red giants
observed with Kepler: comparison of global oscillation parameters from
different methods}. \emph{525}, A131.
\url{https://doi.org/10.1051/0004-6361/201015185}

\leavevmode\hypertarget{ref-huber2011}{}%
Huber, D., Bedding, T. R., Stello, D., Hekker, S., Mathur, S., Mosser,
B., Verner, G. A., Bonanno, A., Buzasi, D. L., Campante, T. L.,
Elsworth, Y. P., Hale, S. J., Kallinger, T., Silva Aguirre, V., Chaplin,
W. J., De Ridder, J., Garc\'ia, R. A., Appourchaux, T., Frandsen, S.,
\ldots{} Smith, J. C. (2011). {Testing Scaling Relations for Solar-like
Oscillations from the Main Sequence to Red Giants Using Kepler Data}.
\emph{743}, 143. \url{https://doi.org/10.1088/0004-637X/743/2/143}

\leavevmode\hypertarget{ref-huber2009}{}%
Huber, D., Stello, D., Bedding, T. R., Chaplin, W. J., Arentoft, T.,
Quirion, P.-O., \& Kjeldsen, H. (2009). {Automated extraction of
oscillation parameters for Kepler observations of solar-type stars}.
\emph{Communications in Asteroseismology}, \emph{160}, 74.
\url{http://arxiv.org/abs/0910.2764}

\leavevmode\hypertarget{ref-matplotlib}{}%
Hunter, J. D. (2007). Matplotlib: A 2D graphics environment.
\emph{Computing in Science \& Engineering}, \emph{9}(3), 90--95.
\url{https://doi.org/10.1109/MCSE.2007.55}

\leavevmode\hypertarget{ref-kjeldsen1995}{}%
Kjeldsen, H., \& Bedding, T. R. (1995). {Amplitudes of stellar
oscillations: the implications for asteroseismology.} \emph{293},
87--106. \url{http://arxiv.org/abs/astro-ph/9403015}

\leavevmode\hypertarget{ref-lightkurve}{}%
Lightkurve Collaboration, Cardoso, J. V. de M., Hedges, C.,
Gully-Santiago, M., Saunders, N., Cody, A. M., Barclay, T., Hall, O.,
Sagear, S., Turtelboom, E., Zhang, J., Tzanidakis, A., Mighell, K.,
Coughlin, J., Bell, K., Berta-Thompson, Z., Williams, P., Dotson, J., \&
Barentsen, G. (2018). \emph{{Lightkurve: Kepler and TESS time series
analysis in Python}} (p. ascl:1812.013).

\leavevmode\hypertarget{ref-mathur2010}{}%
Mathur, S., Garc\'ia, R. A., Régulo, C., Creevey, O. L., Ballot, J.,
Salabert, D., Arentoft, T., Quirion, P.-O., Chaplin, W. J., \& Kjeldsen,
H. (2010). {Determining global parameters of the oscillations of
solar-like stars}. \emph{511}, A46.
\url{https://doi.org/10.1051/0004-6361/200913266}

\leavevmode\hypertarget{ref-mosser2009}{}%
Mosser, B., \& Appourchaux, T. (2009). {On detecting the large
separation in the autocorrelation of stellar oscillation times series}.
\emph{508}(2), 877--887.
\url{https://doi.org/10.1051/0004-6361/200912944}

\leavevmode\hypertarget{ref-nielsen2021}{}%
Nielsen, M. B., Davies, G. R., Ball, W. H., Lyttle, A. J., Li, T., Hall,
O. J., Chaplin, W. J., Gaulme, P., Carboneau, L., Ong, J. M. J., Garc\'ia,
R. A., Mosser, B., Roxburgh, I. W., Corsaro, E., Benomar, O., Moya, A.,
\& Lund, M. N. (2021). {PBjam: A Python Package for Automating
Asteroseismology of Solar-like Oscillators}. \emph{161}(2), 62.
\url{https://doi.org/10.3847/1538-3881/abcd39}

\leavevmode\hypertarget{ref-ricker2015}{}%
Ricker, G. R., Winn, J. N., Vanderspek, R., Latham, D. W., Bakos, G. Á.,
Bean, J. L., Berta-Thompson, Z. K., Brown, T. M., Buchhave, L., Butler,
N. R., Butler, R. P., Chaplin, W. J., Charbonneau, D.,
Christensen-Dalsgaard, J., Clampin, M., Deming, D., Doty, J., De Lee,
N., Dressing, C., \ldots{} Villasenor, J. (2015). {Transiting Exoplanet
Survey Satellite (TESS)}. \emph{Journal of Astronomical Telescopes,
Instruments, and Systems}, \emph{1}(1), 014003.
\url{https://doi.org/10.1117/1.JATIS.1.1.014003}

\leavevmode\hypertarget{ref-schwarz1978}{}%
Schwarz, G. (1978). {Estimating the Dimension of a Model}. \emph{The
Annals of Statistics}, \emph{6}(2), 461--464.
\url{https://doi.org/10.1214/aos/1176344136}

\leavevmode\hypertarget{ref-serenelli2017}{}%
Serenelli, A., Johnson, J., Huber, D., Pinsonneault, M., Ball, W. H.,
Tayar, J., Silva Aguirre, V., Basu, S., Troup, N., Hekker, S.,
Kallinger, T., Stello, D., Davies, G. R., Lund, M. N., Mathur, S.,
Mosser, B., Stassun, K. G., Chaplin, W. J., Elsworth, Y., \ldots{}
Zamora, O. (2017). {The First APOKASC Catalog of Kepler Dwarf and
Subgiant Stars}. \emph{233}(2), 23.
\url{https://doi.org/10.3847/1538-4365/aa97df}

\leavevmode\hypertarget{ref-verner2011}{}%
Verner, G. A., Elsworth, Y., Chaplin, W. J., Campante, T. L., Corsaro,
E., Gaulme, P., Hekker, S., Huber, D., Karoff, C., Mathur, S., Mosser,
B., Appourchaux, T., Ballot, J., Bedding, T. R., Bonanno, A., Broomhall,
A.-M., Garc\'ia, R. A., Handberg, R., New, R., \ldots{} Fanelli, M. N.
(2011). {Global asteroseismic properties of solar-like oscillations
observed by Kepler: a comparison of complementary analysis methods}.
\emph{415}(4), 3539--3551.
\url{https://doi.org/10.1111/j.1365-2966.2011.18968.x}

\leavevmode\hypertarget{ref-scipy}{}%
Virtanen, P., Gommers, R., Oliphant, T. E., Haberland, M., Reddy, T.,
Cournapeau, D., Burovski, E., Peterson, P., Weckesser, W., Bright, J.,
van der Walt, S. J., Brett, M., Wilson, J., Millman, K. J., Mayorov, N.,
Nelson, A. R. J., Jones, E., Kern, R., Larson, E., \ldots{} SciPy 1.0
Contributors. (2020). {{SciPy} 1.0: Fundamental Algorithms for
Scientific Computing in Python}. \emph{Nature Methods}, \emph{17},
261--272. \url{https://doi.org/10.1038/s41592-019-0686-2}

\leavevmode\hypertarget{ref-yu2018}{}%
Yu, J., Huber, D., Bedding, T. R., \& Stello, D. (2018). {Predicting
radial-velocity jitter induced by stellar oscillations based on Kepler
data}. \emph{480}, L48--L53. \url{https://doi.org/10.1093/mnrasl/sly123}

\end{CSLReferences}

\end{document}